\documentclass{aastex}

\usepackage{emulateapj5}

\shorttitle{Size and Structure of Quasar NLRs}
\shortauthors{Bennert et al.}
\begin{document}

\title{Size and Structure of the Narrow--Line Region of Quasars\altaffilmark{1}}

\altaffiltext{1}{Based on observations with the NASA/ESA Hubble 
Space Telescope,
obtained at the Space Telescope Science Institute, which is operated
by AURA, Inc., under NASA contract NAS 5--26555}

\author{Nicola Bennert\altaffilmark{2}, Heino Falcke\altaffilmark{3}, 
Hartmut Schulz\altaffilmark{2}, 
Andrew S. Wilson\altaffilmark{4}, Beverley J. Wills\altaffilmark{5}}

\altaffiltext{2}{Astronomisches Institut Ruhr--Universit\"{a}t Bochum, 
           Universit\"{a}tsstra{\ss}e 150, D--44780 Bochum, Germany; email:
nbennert@astro.ruhr--uni--bochum.de; email: hschulz@astro.ruhr--uni--bochum.de}

\altaffiltext{3}{Max--Planck--Institut f\"{u}r Radioastronomie, 
Auf dem H\"{u}gel 69, D--53121 Bonn, Germany; email: 
hfalcke@mpifr--bonn.mpg.de}

\altaffiltext{4}{Astronomy Department, University of Maryland, College Park, 
MD 20742--2421, USA; email:wilson@astro.umd.edu}

\altaffiltext{5}{Department of Astronomy and McDonald Observatory, 
University of Texas at Austin, Austin, TX 78712, USA; email:
bev@pan.as.utexas.edu}


\begin{abstract}
We have observed the narrow--line regions (NLRs) of the seven 
brightest radio--quiet PG (or BQS) quasars (z $<$ 0.5) with the Wide Field
and Planetary Camera 2 on board the Hubble Space Telescope (HST). Linear--ramp
filters were used to image the [\ion{O}{3}]\,$\lambda$5007 line
emission with 0\farcs0455--0\farcs1 pixel resolution. We find that the NLRs are
very compact with typical extents of 2\arcsec--4\arcsec. Two quasars
show compact filamentary structures similar to Seyfert NLRs.
They may be related to radio outflows.  Most interestingly, when
including a sample of Seyfert galaxies observed with HST, we
tentatively find that the size of the NLR is proportional to 
the square root of the [\ion{O}{3}] luminosity. This is comparable to
the scaling found for the size of the broad--line region with
continuum luminosity, which has been interpreted in terms of a constant
photoionization parameter. The relation determined here connects the
NLR of radio--quiet quasars and Seyferts over three orders of
magnitude in [\ion{O}{3}] luminosity.
\end{abstract}



\section{INTRODUCTION}
Quasars are active galactic nuclei (AGN) 
in which 
two different regions of ionized gas can be distinguished --- the
broad--line region (BLR) and the narrow--line region (NLR) which
exhibit optical emission lines, kinematically broadened with typical
widths of 10$^{3-4}$\,km\,s$^{-1}$ and 10$^{2-3}$\,km\,s$^{-1}$,
respectively.  The optical spectra of quasars resemble those of the
less luminous Seyfert galaxies, and it is presumed that most
radio--quiet quasars are their more luminous counterparts. 
The most
prominent optical emission lines of the NLR are 
[\ion{O}{3}]\,$\lambda$5007 (hereafter referred to simply as
[\ion{O}{3}]) and H$\alpha$\,+\,[\ion{N}{2}]
$\lambda\lambda$6548,6584.

In the Unified Model of AGN, an optically thick obscuring dust torus
is envisioned to encircle the accretion disk
\citep{ant93},
leading to a so--called ``ionization cone'' formed by anisotropic escape
of ionizing photons \citep{sch88,pog88,pog89,sto92}. 
Thus, the NLR is expected to show a specific
morphology, which is suited for investigation with the high spatial
resolution afforded by the Hubble Space
Telescope (HST). Such investigations have been carried out for
Seyfert galaxies, revealing highly elongated structures
or ionization cones \citep{cap96,fal98}. 

While the NLR in Seyferts is now relatively well studied, there are no
comparable studies for the NLR in 
quasars.
By observing quasars we can greatly extend the luminosity range to answer
questions such as:
Are quasars indeed just
scaled--up versions of Seyferts?  Does the size of the NLR scale
with luminosity? 
Is this emission--line region possibly affected by radio
jets shaping the interstellar medium, 
or is there a contribution from star formation?

We make a first step in this direction by presenting an
HST emission--line imaging survey of a complete sample of
the seven brightest [in [\ion{O}{3}], \citet{bor92}]
radio--quiet quasars from the BQS \citep{sch83,kel89}
with z $<$ 0.5. The sample is 
given in Table~\ref{tabobs}. Here we present an
investigation of the NLR structure as seen in the [\ion{O}{3}] line
and discuss the size of the NLR as a function of luminosity.

Luminosity distances for both Seyferts and quasars were
calculated by using redshifts relative to the 3K background as derived
with the velocity calculator provided by the NASA Extragalactic
Database (NED).  
Throughout this paper we adopt a Hubble constant of
$H_0$ = 65 km\,s$^{-1}$\, Mpc$^{-1}$ and a homogeneous, isotropic,
flat world model, which includes Einstein's cosmological constant
$\Lambda$ in agreement with recent supernova measurements \citep{per99}: 
$\Sigma \Omega$ = 1, $\Omega_{{\rm matter}}$ = 0.35,
$\Omega_{{\rm radiation}}$ = 0.05, and $\Omega_{\Lambda}$ = 0.6.

\section{OBSERVATION, DATA REDUCTION, \& ANALYSIS}

\subsection{HST Observations}
All quasars in our sample were observed with the Wide Field
and Planetary Camera 2 (WFPC2) on board HST
between January and October 2000.  Except for observations of the
adjacent continua, we used the linear--ramp filters (LRFs).  A description
of properties and data reduction of LRF data is given in \citet{fal98}.

For our observations, 
the filters and central wavelengths were chosen
to be centered on the redshifted [\ion{O}{3}] emission
lines. \objectname{PG0026+129} was the only quasar
that could be imaged with the Planetary Camera and its smaller pixel scale
of 0\farcs0455 pixel$^{-1}$. The other quasar images were
taken with the Wide Field Camera (0\farcs1 pixel$^{-1}$).
Several short integrations of a bright star (V = 11--16)
were taken in the LRF at the same position on the chips as the
corresponding quasar, to determine the point--spread--function (PSF) for
calibration purposes.

To subtract the underlying continuum, images of the adjacent continua
were taken in narrow--band filters (F588N: $\delta \lambda = 49$\AA, F631N:
 $\delta \lambda = 30.9$\AA, and F673N:  $\delta \lambda = 47.2$\AA),
at the same chip
position as the LRF images.  All observations were split into several
integrations to allow for cosmic--ray rejection with
integration times chosen to avoid saturation at 
the position of the central point source.

\subsection{Data Reduction \& Analysis}
All exposures were bias-- and dark--subtracted
and, except for the LRF images,
flat--field corrected by WFPC2 pipeline
processing at the Space Telescope Science Institute (STScI). 
For the LRFs, we used
flat fields taken in nearby narrow--band filters.
In all cases the
different exposures were shifted (using linear interpolation as
recommended for subsampled data) by aligning the central intensity
peaks of the LRF, continuum, and PSF images, as determined from 
gaussian--profile fitting.
The point--source calculator (as provided by STScI) 
was used for flux calibration.  For subtraction
of the continuum and especially the contribution from the unresolved
continuum source at the center of the quasar, we used either the
continuum images or the PSF star (see Table 1).
In one case
(\objectname{PG1012+008}), no continuum was detected on the continuum
images, and a scaled PSF star was substracted from the LRF image. 
The scaling was done such that, after subtraction, the
central pixel had the same value as the average of the neighbouring
pixels. This avoids ``holes'' in the center and may lead to an
underestimate of the [\ion{O}{3}] emission.
We also had to scale the ``continuum'' image for
\objectname{PG1049--005} in the same way,
as the image was, by accident, not taken in a continuum
wavelength range 
but in the middle of the emission line of [\ion{O}{3}]\,$\lambda$4959.
For \objectname{PG0157+001} and \objectname{PG0953+414}
the continua were clearly extended compared with that of the PSF star,
and the continuum images were used for subtraction.
In the three remaining cases
(\objectname{PG0026+129}, \objectname{PG0052+251}, and
\objectname{PG1307+085}) the continua were not extended, compared with
the PSF star, and we subtracted the PSF star, scaled to the flux seen
in the continuum image, because of the higher signal--to--noise ratio
of the star.

To check how much galaxy line emission or 
continuum we might be missing by using
our short--exposure continuum (or scaled PSF) images, we used broad--band
continuum images of the host galaxies of our quasars from
\citet{bah97}. The broad--band images were scaled to the appropriate filter
width and exposure time of our LRF images and subtracted. We found the
same [\ion{O}{3}] flux at radii $>2-10$ pixels, where the broad--band images
were not overexposed, and concluded that, indeed, the host--galaxy
continuum contribution to the LRF images was negligible outside the
very nucleus. 
We also performed surface photometry of the continuum--subtracted
LRF images and compared the resulting [\ion{O}{3}]
luminosities (Table~\ref{tabflux}) with the 
[\ion{O}{3}] magnitudes determined by \citet{bor92}
with ground--based spectroscopy in a 6\arcsec $\times$ 1\farcs5
extraction aperture. After correcting the luminosity distances
for differences in the Hubble constant (a 30\% effect) and
the world model ($\sim$ 3\%),
the difference between the HST imaging and
ground--based spectroscopic data scatters within 0.25 mag, consistent
with the expected spectrophotometric errors.
The evidence suggests that we have not missed any significant extended
flux, and that our continuum subtraction is correct.

\section{RESULTS}
\subsection{Photometry \& Structure}
For each quasar we defined a radius of the 
NLR in our images at which a 3 $\sigma$--level
above background was reached in ring apertures. The surface brightness
at that distance was typically 0.3\% of
central, thereby enclosing
more than 98\% of the detectable emission. 
The results of the photometry as well as these
radii are given in Table~\ref{tabflux}. 
In all cases the NLR was largely concentrated within
2\arcsec--4\arcsec.

\begin{figure}
\includegraphics[width=8cm]{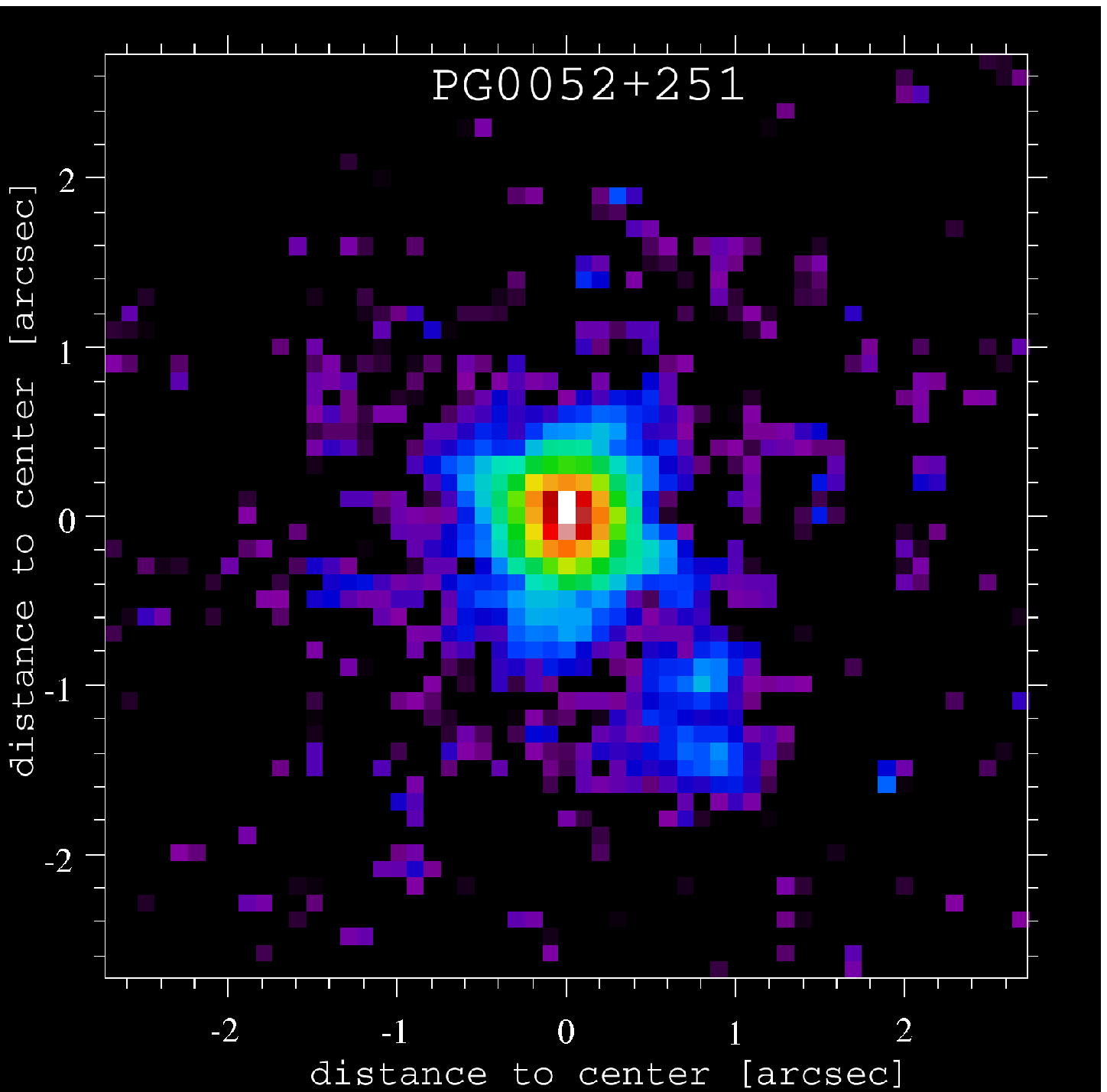}
\includegraphics[width=8cm,clip]{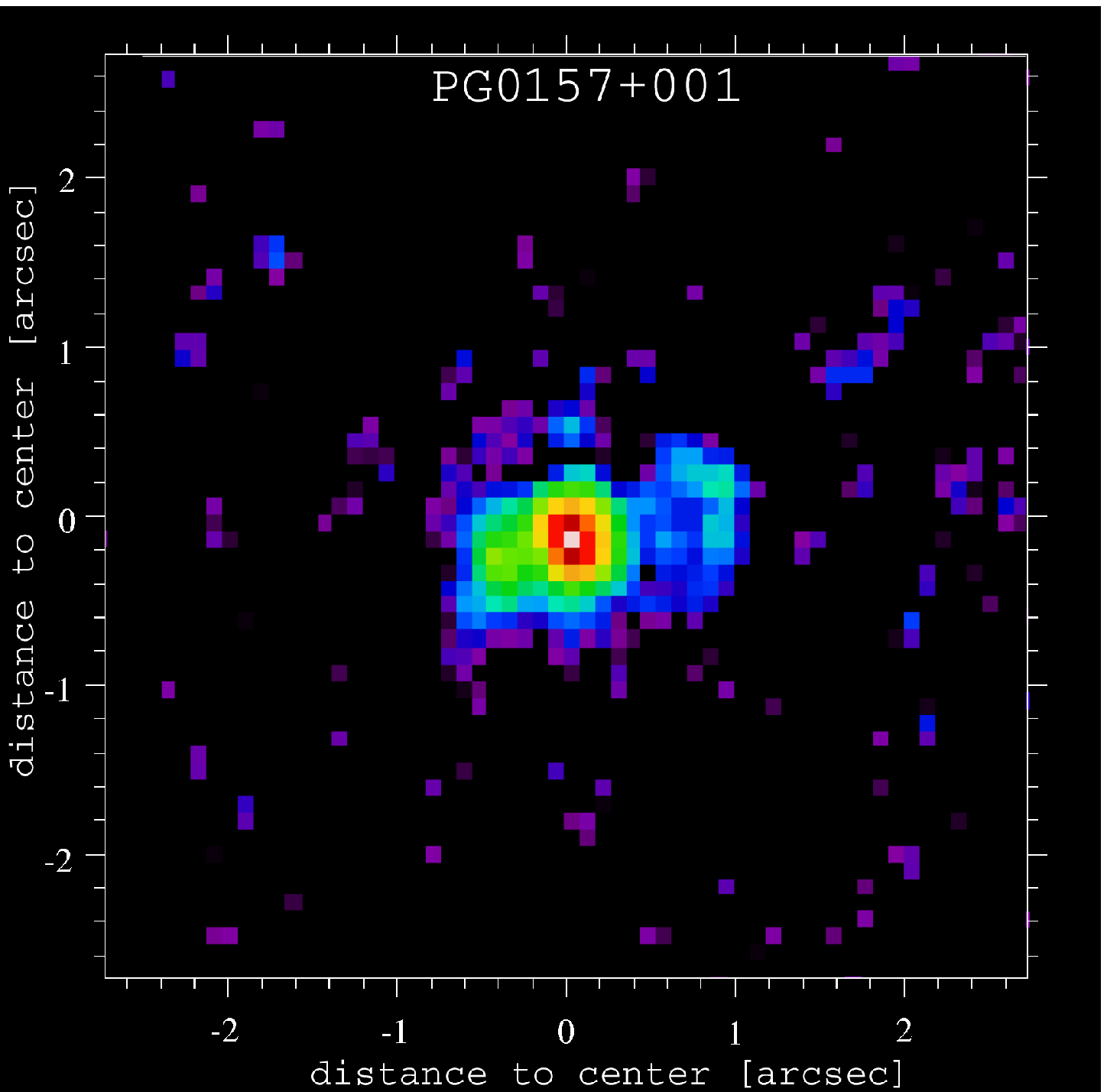}
\caption{
LRF images of PG0052+251 (left) and PG0157+001 
(right)
in [\ion{O}{3}]\,$\lambda$5007 in
a logarithmic scale in false colours. North is
up, east is to the left. The tick marks are chosen such 
that the central pixel is at 0\farcs0; major tick marks
are in distances of one arcsecond.}
\label{feat}
\end{figure}

Two quasars
show compact filamentary
structure as shown in Figure~\ref{feat}.  Photometry of
these features (\objectname{PG0052+251}: $\sim$ 1\farcs5 to the south and
$\sim$
0\farcs8 to the west of the nucleus; \objectname{PG0157+001}: 
$\sim$ 0\farcs4 to
the north and $\sim$ 1\arcsec~to the west of the nucleus), was carried
out separately (see Table~\ref{tabflux}). The feature in 
\objectname{PG0052+251} can
also be found in the broad--band images of \citet{bah97} as ``knots''
and in a radio map from \citet{kuk98} (classified as possible core--jet), 
whereas \citet{mil93} do not find a significant
radio extension.
The ``linear'' structure of \objectname{PG0157+001} 
is coincident with radio features 
[\citet{kuk98} (classified as a double source
with radio components on either side of the optical nucleus), \citet{mil93}].
A close association between the NLR emission--line morphology 
and that of the radio emission has already been found
for Seyferts \citep{cap96,fal98}.
\objectname{PG1012+008} and \objectname{PG0026+129}
reveal some extended radio structure \citep{mil93,kuk98},
but we do not find extended emission--line features.  
Five out of the seven quasars are listed
in \citet{kuk98} with maximum radio angular extents of
0\farcs24--2\arcsec.

\subsection{Size as a Function of Luminosity}
An important question
is how the NLR
grows as one increases the luminosity of the central engine from Seyfert
to quasar luminosities. 
For that reason we compared the sizes and luminosities of the NLR in
our quasar sample with the NLR of a sample of Seyfert 2 galaxies
that we have studied earlier. This sample was observed with HST in a
very similar fashion, using LRFs and imaging the [\ion{O}{3}] emission
line \citep{fal98}. In Figure~\ref{o3corr} we plot the
linear sizes versus the 
luminosities for the Seyferts and quasars.

The number of sources is of course very limited, but with the addition
of the quasars we are probing three orders of magnitude in luminosity.
The weighted linear least--square fit is 
(weight = $\sigma^{-2}$,
$\sigma$ = standard error in $\log R$;
units of $R$ and $L$ are pc and erg/s, respectively): 
\begin{equation}
\log(R_{\rm {NLR}}) = (0.52 \pm 0.06) \log(L_{\rm [OIII]}) - (18.5 \pm 2.6)
\end{equation}
with a correlation coefficient (cc) of 0.92 for 14 datapoints (dp).  Hence, the
size of the NLR seems to scale proportionally with the square
root of the [\ion{O}{3}] luminosity.

We note that for the quasars alone a similar relation is found, when
plotting the NLR size versus the H$\beta$ luminosity (Figure~\ref{o3corr}, 
right).
With $L_{{\rm H}\beta}$ derived by multiplying our $L_{\rm [OIII]}$
with the H$\beta$/[\ion{O}{3}] ratios from \citet{bor92} 
(see Table~\ref{tabflux})
we obtain (cc = 0.85, 7 dp)
\begin{equation}
\log(R_{\rm {NLR}}) = (0.67 \pm 0.15) \log(L_{\rm {H}\beta}) 
- (25.5 \pm 6.3) .
\end{equation}

The average [\ion{O}{3}]/H$\beta$ ratio
of the Seyferts is 30 times larger than
that of the quasars.
Since the Seyferts are Type 2 objects,
in the framework of the unified model,
only a few percent of the total H$\beta$ is visible.
\citet{bor92} also found that $\sim$ 3\% 
of the total H$\beta$ flux is visible as the narrow line component.
We have thus multiplied the inferred narrow line 
H$\beta$ luminosities of the Seyferts by a factor 
of 30 to estimate their total H$\beta$ luminosities,
and found that the resulting $R$ versus $L_{H\beta}$
relationship
is consistent with an extrapolation to lower luminosity of the relationship 
found for the quasars.
Including both Seyfert and quasar data into a weighted linear least--square
fit we obtain (cc = 0.91, 14 dp)
\begin{equation}
\log(R_{\rm {NLR}}) = (0.49 \pm 0.06) \log(L_{\rm {H}\beta})
- (17.5 \pm 3.4) .
\end{equation}

\section{DISCUSSION \& CONCLUSIONS}

We find that the NLRs of seven bright radio--quiet PG quasars
are remarkably compact with typical extents of
2\arcsec--4\arcsec. Hence, detailed imaging of quasar NLRs requires
sub--arcsecond resolution. Generally, the structure is relatively
symmetric, in agreement with the
unified scheme, that predicts a view 
into the ionization cones of these type 1 objects.

Two quasars exhibit compact filamentary
structure like that seen for Seyferts.
These structures may be related to radio
outflows. This is reminiscent of the situation in many Seyfert galaxies,
where radio outflows are morphologically related to the NLR.  
In three quasars the radio emission is approximately as compact as the
NLR, and in all five cases that can be found in literature,
the maximum angular extent is in agreement with our compact
extents. 

The NLR in quasars seems to be consistent with being a
scaled--up version of the NLR in Seyferts. In fact, one of the Seyfert 2
galaxies, \objectname{Mrk 34}, has a similar radius and [OIII]
luminosity to the quasars, and hence might be
considered a ``type 2 quasar''. In addition, the size of the
NLR seems to scale roughly with the square root of the [\ion{O}{3}]
luminosity when combining the quasar and Seyfert samples.

The latter result, however, has to be taken with some caution. For
one, we have a very limited number of sources available. Especially
for quasars, high resolution images of the NLR are extremely
scarce, one essentially needs the resolution of HST plus the
flexibility of the LRFs to image redshifted emission lines. Second,
when comparing the Seyfert 2 and the quasar sample, one needs to
consider that, according to the unified scheme, orientation effects
might bias the type 2 sample towards somewhat larger sizes. Third,
the individual scatter of NLR sizes is relatively large. Fourth, the
``size'' of the NLR is not a well defined quantity, being dependent on
sensitivity and resolution. Nevertheless,
given the large span in luminosities, none of the effects is likely
to change the results significantly.

Seeking a clue to the origin of this
relation, we consider a recombination line like
H$\beta$, rather than [\ion{O}{3}], as a tracer of continuum luminosity. 
The ionization parameter is given by $U =
Q/(4 \pi c n_e R_u^2)$ ($Q =$ rate of H--ionizing photons, $R_u$ =
distance between photoionizing source and emission--line clouds).
Employing the relation $\omega Q =
\alpha_B/(\alpha_{{\rm H}\beta} h \nu_{{\rm H}\beta}) = 3.9\,10^{12}
L_{{\rm H}\beta}$ (cgs units; $\omega$ is the covering factor and the
recombination coefficients $\alpha_B$ and $\alpha_{{\rm H}\beta}$ are
taken from \citet{ost89} for $T=2\,10^4$K) yields a relation $R_u
\propto L_{{\rm H}\beta}^{0.5}$ for given $U$, $n_e$, and constant
$\omega$.

\begin{figure}
\epsscale{1.1}
\plottwo{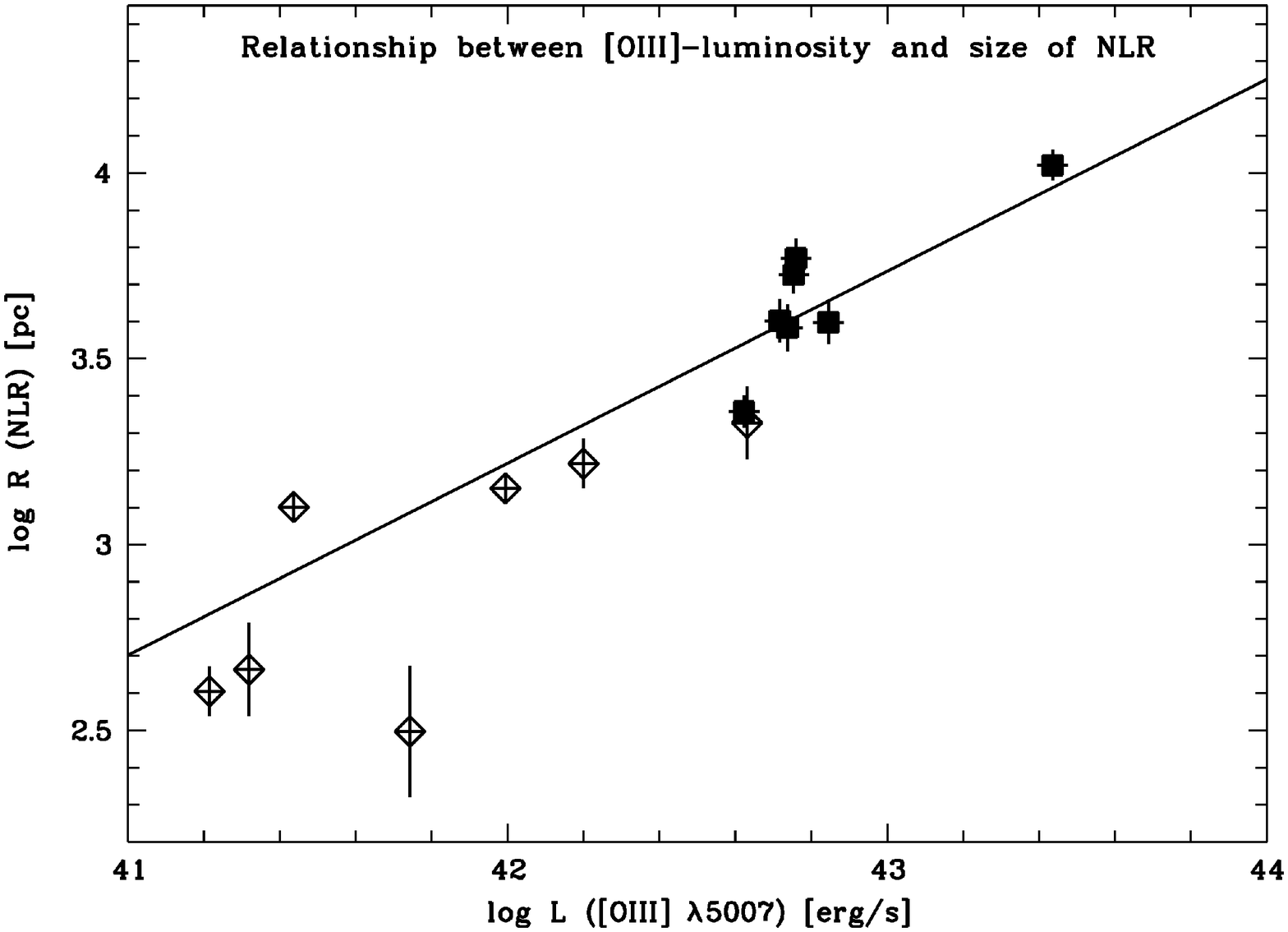}{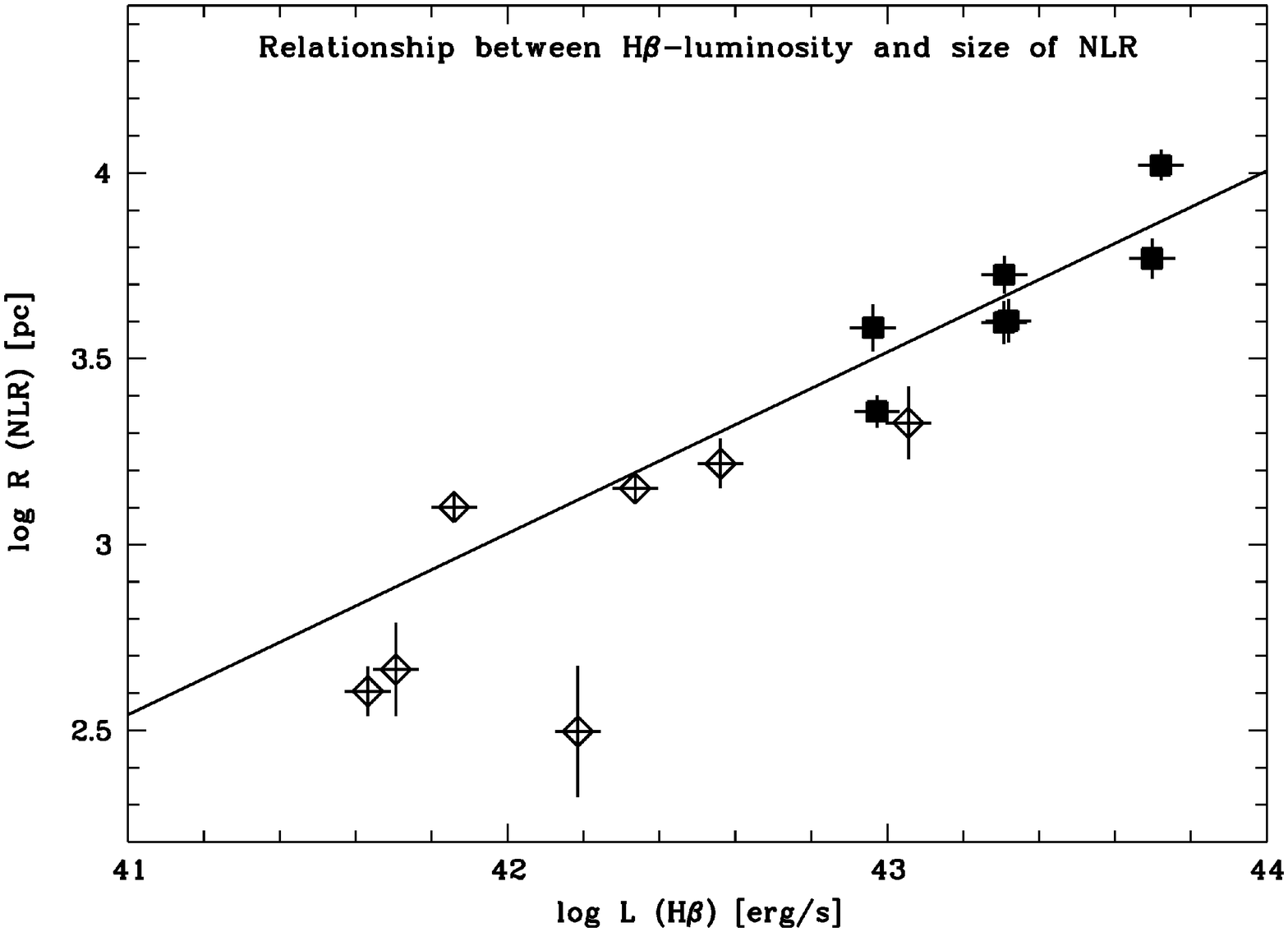}
\caption{These panels show the radius of the NLR
versus the emission--line luminosity in [\ion{O}{3}]\,$\lambda$5007 (left)
and in H$\beta$ (right) 
on logarithmic scales.
The open diamonds are Seyferts, the filled 
squares are PG quasars. The solid lines show
weighted linear
fits resulting in R $\propto$ ${L_{\rm [OIII]}}^{0.52}$ 
(left hand panel) and
R $\propto$ ${L_{\rm H\beta}}^{0.49}$ (right).
The error bars indicate
the uncertainties in the fluxes from both the photometry and the placement
of the apertures.
}\label{o3corr}
\end{figure}

The slope in
Figure~\ref{o3corr} is close to the slope of $0.6\pm0.1$~given by
\citet{pet01} for the relationship between BLR size 
(measured from reverberation mapping) and the
continuum--luminosity (at 5100\AA). \citet{mcl02}
find an even tighter correlation between BLR radius and 3000\AA~luminosity, 
with both correlations being consistent with a
relation of the form $R_{\rm {BLR}}\propto L^{0.5}$.

To first order, BLRs and NLRs exhibit similar ionization parameters
commonly explained by interaction of clouds with a quasar wind \citep{sch86}. 
More recently,
\citet{dop02} use radiation--pressure dominated
photoionization models to explain a constant ionization parameter (in
the range of --2.5 $\le$ $\log U$ $\le$ --2).  At the outskirts of a NLR,
$n_e\sim 10^{2-3}$ cm$^{-3}$, $U\sim 10^{-(2-3)}$, so that efficient
[\ion{O}{3}] emission comes from regions with $U n_e \approx 1$,
corresponding to an ionizing photon rate $Q/(4 \pi R_u^2) = 3~10^{10}$
photons/(s cm$^2$).  Individual values of $R_u$ derived with this
scheme are slightly larger than the measured $R$, corresponding to
covering fractions $\omega$ $\le$ 1.
  
The same average [\ion{O}{3}]/H$\beta$
ratio for Seyferts and quasars preserves the
slope of the size--[\ion{O}{3}] relation for a size--H$\beta$
relation. 
This corroborates the explanation in terms of the
ionization parameter and argues for photoionization as
the main underlying process for producing narrow lines.

As pointed out by the referee, the relationship
$L_{\rm [OIII]}$ (or $L_{H\beta}$) $\propto R_{\rm NLR}^2$ 
corresponds to constant [\ion{O}{3}] (or 
H$\beta$) surface brightness. We are, however, not aware of any other 
physical, or instrumental, reason why this should be the case,
since [\ion{O}{3}] is emitted
by the whole {\em volume} filled with NLR clouds.

Clearly, more HST observations of quasars with lower [\ion{O}{3}]
luminosity are needed to test the validity of the correlation.  We
expect fainter [\ion{O}{3}] quasars to have even smaller NLRs than the
sources presented here and such a program will present an
observational challenge for the years to come. Nevertheless, the
available data provides the first direct evidence that quasar and
Seyfert NLR are related and possibly evolve along a common
luminosity--size track determined by photoionization.

\acknowledgments
N.B. thanks T. J\"{u}rges for computer
support, C. Leipski for discussions, and
R. Chini for financial support.
This research was also supported by STScI 
through grant GO 8239 to the Universities of Maryland and Texas.



\begin{deluxetable}{lccccccc}
\tabletypesize{\small}
\tablecolumns{8}
\tablewidth{0pc}
\tablecaption{Details of Observation and Reduction}
\tablehead{
\colhead{Quasar} & \colhead{$\alpha$} & \colhead{$\delta$} & 
\colhead{z} & \colhead{Int. [\ion{O}{3}]} & \colhead{Int.
 Cont.}
& \colhead{Int. PSF Star} & \colhead{Subtr. Method} \\
\colhead{(1)} & \colhead{(2)} & \colhead{(3)} & \colhead{(4)} & \colhead{(5)}
& \colhead{(6)} & \colhead{(7)} & \colhead{(8)}
}
\startdata
\objectname{PG0026+129} & 00 29 13.71 & +13 16 03.83 & 0.142 &
300 (8) & 120 (2) & 6 (2) & PSF star scaled to cont.\\
\objectname{PG0052+251} & 00 54 52.13 & +25 25 39.33 & 0.155 &
280 (5) & 60 (4) & 35 (4) & PSF star scaled to cont.\\
\objectname{PG0157+001} & 01 59 50.21 & +00 23 41.55 & 0.163 &
240 (6) & 60 (4) & 40 (4) & cont.\\
\objectname{PG0953+414} & 09 56 52.35 & +41 15 22.53 & 0.234 &
165 (8) & 60 (4) & 50 (4) & cont.\\ 
\objectname{PG1012+008} & 10 14 58.10 & +00 32 05.44 & 0.187 & 
400 (4) & \nodata & 60 (3) & scaled PSF star \\
\objectname{PG1049--005} & 10 51 51.46 & --00 51 18.16 & 0.360 &
212.5 (8) & 60 (4) & 40 (4) & scaled cont.\\
\objectname{PG1307+085} & 13 09 47.93 & +08 19 49.62 & 0.155 &
250 (6) & 60 (4) & 35 (4) & PSF star scaled to cont.\\
\enddata
\tablecomments{Columns: (1) quasar; (2--3)  right ascension 
(in units of hours, minutes, and seconds) and declination
(degrees, arcminutes, and arcseconds)
as provided by NED (J2000); (4) heliocentric redshift from NED; (5--7) 
mean of the integration time in s for 
the single image ([\ion{O}{3}], continuum or PSF star) followed by,
in brackets, the number of exposures which were combined; 
(8) subtraction method: scaled PSF star,
continuum, scaled continuum or PSF star weighted by the flux in the continuum
image}
\label{tabobs}
\end{deluxetable}


\begin{deluxetable}{lccccccc}
\tabletypesize{\small}
\tablecolumns{8}
\tablewidth{0pc}
\tablecaption{Emission--line flux and size of the NLRs}
\tablehead{
\colhead{Quasar} & \colhead{Lum. Dist.} & \colhead{Apert.} 
& \colhead{Radius} 
& \colhead{[\ion{O}{3}] Flux} 
& \colhead{[\ion{O}{3}] Lum.} & 
\colhead{H$\beta$ Lum.} & \colhead{Surf. Brightn.}\\
\colhead{(1)} & \colhead{(2)} & \colhead{(3)}  & \colhead{(4)} & \colhead{(5)}
& \colhead{(6)} & \colhead{(7)} & \colhead{(8)}}
\startdata
\objectname{PG0026+129} & 708 & 0.86 & 2281 $\pm$ 240 & 7 & 4.2 $\pm$ 0.4 & 9.41 $\pm$ 1.4 & 1.47\\
\objectname{PG0052+251} & 780 & 1.39 & 3958 $\pm$ 565 
& 8.59 & 6.25 $\pm$ 0.6 & 18.12 $\pm$ 2.7 & 2.03\\
(feature) & & 0.6 & 1696 $\pm$ 283 & 0.38 & 0.27 $\pm$ 0.03 & 0.81 $\pm$ 0.1\\ 
\objectname{PG0157+001} & 824 & 1.29 & 3832 $\pm$ 
590 & 5.67 & 4.61 $\pm$ 0.5 & 7.74 $\pm$ 1.2 & 2.43\\
(feature) & & 0.5 & 1474 $\pm$ 295 & 0.4 & 0.32 $\pm$ 0.03 & 0.54 $\pm$ 0.1\\
\objectname{PG0953+414} & 1239 & 1.49 & 5884 $\pm$ 785 & 3.14 & 5.76 $\pm$ 0.6 
& 49.9 $\pm$ 7.5 & 1.05\\
\objectname{PG1012+008} & 973 & 1.59 & 5319 $\pm$ 665 & 5.01 & 5.67 $\pm$ 0.6 & 20.35 $\pm$ 3.1 & 1.54\\
\objectname{PG1049--005} & 2011 & 1.99 & 10484 $\pm$ 1048 & 5.64 & 27.28 $\pm$ 2.7 & 52.57 $\pm$ 7.9 & 0.53\\
\objectname{PG1307+085} & 791 & 1.39 & 4002 $\pm$ 572 & 6.96 & 5.21 $\pm$ 0.5 & 20.83 $\pm$ 3.1 & 1.35\\
\enddata
\tablecomments{Columns: (1) quasar; (2) luminosity distance 
in Mpc; (3) radius of the apertures in arcseconds; 
(4) radius of the NLR in pc; 
(5) [\ion{O}{3}]\,$\lambda$5007 fluxes, 
given in units of $10^{-14}$ \,erg\,s$^{-1}$\,cm$^{-2}$. These
fluxes were derived by summing the flux in the HST images
inside the circular aperture with radius given in column (3);
(6) [\ion{O}{3}]\,$\lambda$5007 luminosities in units of $10^{42}$\,erg\,s$^{-1}$; (7) H$\beta$ luminosities in units of $10^{42}$\,erg\,s$^{-1}$; 
(8) surface brightness
at the edge of the NLR in units of
$10^{-17}$\,erg\,s$^{-1}$\,pixel$^{-1}$ 
(those of the Seyferts
were at about the same value)}
\label{tabflux}
\end{deluxetable}


\end{document}